\documentclass[epj]{svjour}
\usepackage{graphicx}
\usepackage{amssymb,amsmath}
\begin{document}
\title{Direct-Photon Production from SPS to RHIC Energies}
\author{Klaus Reygers\inst{1} for the PHENIX collaboration
}                     
%
%
\institute{Institut f\"{u}r Kernphysik, University of M\"{u}nster, Germany}
\date{Received: date / Revised version: date}
%
\abstract{ 
  Direct photons are an important tool for the detection of the
  quark-gluon plasma in ultra-relativistic nucleus-nucleus collisions.
  Direct-photon measurements were made in Pb+Pb collisions at
  $\sqrt{s_\mathrm{NN}}=17.2\,$GeV and in Au+Au collisions at
  $\sqrt{s_\mathrm{NN}}=200\,$GeV. These results are reviewed and
  compared with model calculations.
  \PACS{ {PACS-key}{describing text of that key} \and
    {PACS-key}{describing text of that key}
     } 
} 
\maketitle
\section{Introduction}
\label{sec:intro}
Measurements of direct photons, {\it i.e.} of photons which are not
produced in electromagnetic decays like $\pi^0 \rightarrow
\gamma+\gamma$ and $\eta \rightarrow \gamma+\gamma$, have a long
history \cite{Ferbel:1984ef,Peitzmann:2001mz}.  First measurements of
direct photons in proton-proton collisions were made in the late
1970's and hinted at the presence of point-like charged objects within
the proton.  Quantum chromodynamics (QCD) was developed at this time
and made predictions for direct-photon production. The experimental
test of QCD became the primary motivation for direct-photon
measurements. QCD is now considered the correct theory of the strong
interaction and the focus of direct-photon measurements in collisions
of hadrons has shifted towards constraining the gluon distribution
function within hadrons. Direct photons are specially suited for this
because gluons are involved at leading order in photon production via
the quark-gluon Compton scattering ($\mathrm{q}+\mathrm{g} \rightarrow
\mathrm{q} + \gamma$) whereas gluons are involved only at
next-to-leading order in processes like deep inelastic scattering of
leptons.

The interest in direct photons in collisions of heavy nuclei arises
from the fact that once they are produced photons leave the hot and
dense fireball virtually without further interaction. They can thus
convey information about the early stage of a nucleus-nucleus
collision. It is expected that in the early stage of a nucleus-nucleus
collision with sufficiently high energy a thermalized medium is formed
in which quarks and gluons are the relevant degrees of freedom.  The
electric charges in such a quark-gluon plasma (QGP) are expected to
radiate photons whose momentum distribution reflects the temperature
of the system. The production of thermal photons was suggested as a
QGP signature \cite{Feinberg:1976ua,Shuryak:1978ij}. The fireball
produced in a nucleus-nucleus collision expands and cools.  The
dominant contribution to the spectrum of thermal direct photons is
expected from the early hot phase directly after thermalization. Thus,
by measuring thermal direct photons one can constrain the initial
temperature of the fireball.

\begin{figure}
  \centerline{
    \includegraphics[width=0.45\textwidth]{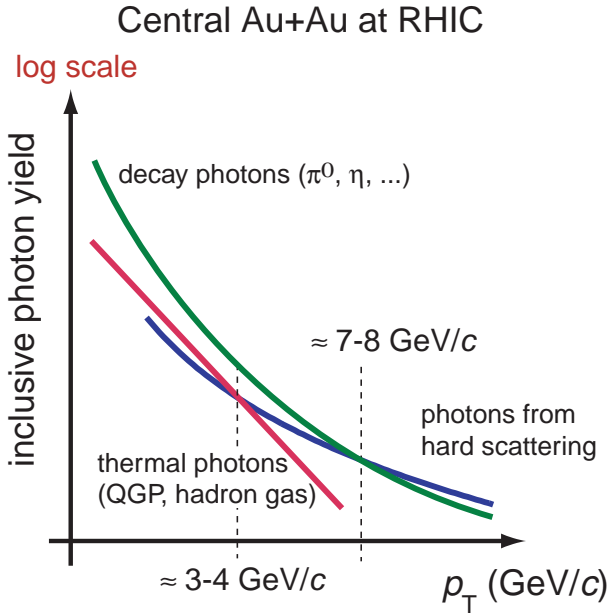}}
  \caption{Schematic representation of the different contributions to
    the inclusive photon spectrum in Au+Au collisions at RHIC.  Over a
    wide transverse momentum range thermal direct photons from the QGP
    and the hadron gas as well as prompt photons from initial hard
    parton scattering are overwhelmed by photons from hadron decays.
    The strong suppression of neutral pions and $\eta$ mesons in
    central Au+Au collisions at RHIC ($\sqrt{s_\mathrm{NN}} =
    200\,$GeV) results in a reduced decay photon background and
    eases the extraction of a direct-photon signal.}
  \label{fig:cartoon} 
\end{figure}

The transverse momentum spectrum of thermal photons is expected to
decrease approximately as $\exp(-p_\mathrm{T}/T)$ for temperature $T$.
Direct photons from initial parton-parton scatterings with high
momentum transfer (hard scattering) follow a power-law shape
$p_\mathrm{T}^{-n}$ and thus dominate the direct-photon spectrum at
sufficiently high $p_\mathrm{T}$. This is schematically shown in
Figure~\ref{fig:cartoon}. The measurement of direct photons at high
$p_\mathrm{T}$ in nucleus-nucleus collisions allows to test the
expected scaling of particle yields in hard processes with the number
of inelastic nucleon-nucleon collisions ($N_\mathrm{coll}$). This is
of paramount importance for the interpretation of the suppression of
hadrons at high-$p_\mathrm{T}$ in central Au+Au collisions at RHIC as
a final state effect, {\it i.e.}  as an effect that is caused by the
medium created in those collisions.  A common interpretation is that
high-$p_\mathrm{T}$ hadron suppression is the result of energy loss of
partons in the medium after their initial hard scattering (jet
quenching) \cite{Vitev:2002pf}.

\section{Production Mechanisms}
\label{sec:prod}
For the production of direct photons in collisions of hadrons complete
next-to-leading order (NLO) calculations in perturbative QCD (pQCD)
are available. In these calculations the total direct-photon cross
section can be considered as the sum of a direct component and a
bremsstrahlung component \cite{Aurenche:1998gv}:
\begin{equation}
  \sigma_\mathrm{tot}^{\mathrm{direct} \gamma}
  = \sigma_\mathrm{direct} + \sigma_\mathrm{brems}.
\end{equation}
The dominant contributions to the direct component come from
quark-antiquark annihilation ($\mathrm{q}+\bar{\mathrm{q}} \rightarrow
\mathrm{g} + \gamma$) and quark-gluon Compton scattering
($\mathrm{q}+\mathrm{g} \rightarrow \mathrm{q} + \gamma$).  The
bremsstrahlung component contains calculable processes in which a
photon is radiated off a quark (\textit{e.g.}  $\mathrm{q}+\mathrm{g}
\rightarrow \mathrm{q}+\mathrm{g} +\gamma$) but also photons produced
in the fragmentation of quarks and gluons into hadrons. The latter
process is a soft process that cannot be described by pQCD and is
therefore described by a phenomenological parton-to-photon
fragmentation function.

An intrinsic uncertainty in the pQCD calculation arises due to the
choice of unphysical and arbitrary renormalization, factorization, and
fragmentation scales. The resulting uncertainties of the calculated
direct-photon transverse momentum ($p_\mathrm{T}$) spectra are
typically on the order of 20-30\% \cite{Aurenche:1998gv}. The
comparison of data and pQCD calculations in
$\mathrm{p}+\mathrm{p}(\bar{\mathrm{p}})$ collisions over a wide range
of collision energies ($\sqrt{s}$) shows a systematic pattern of
deviation: the measured direct-photon $p_\mathrm{T}$ distributions are
typically steeper than the QCD prediction and so in many comparisons
the measured direct-photon yields at low-$p_\mathrm{T}$ lie above the
QCD expectations \cite{Huston:1995vb}. With a non-perturbative
parameter $k_\mathrm{T}$, which reflects an initial state transverse
momentum component of the partons prior to the hard scattering, it is
possible to improve the agreement between data and QCD.

In ultra-relativistic collisions of nuclei additional direct-photon
sources are expected. One generally distinguishes between thermal and
non-thermal direct photons. Non-thermal direct photons include prompt
(or pQCD) photons, which result from hard initial parton-parton
scattering analogous to the production mechanism in
$\mathrm{p}+\mathrm{p}(\bar{\mathrm{p}})$ collisions. Due to intense
rescattering among the quarks and gluons produced in a nucleus-nucleus
collision one expects that a thermalized QGP is formed after a
formation time $\tau_0$. Pre-equilibrium direct photons produced after
the initial parton-parton scattering but before thermalization are a
second source of non-thermal photons. So far the theoretical
description of pre-equilibrium direct-photon production is not well
under control. The QGP expands and cools and at temperature of
$T_\mathrm{c} \approx 170$~MeV a transition to a gas of hadrons is
expected \cite{Karsch:2003jg}.  Thermal direct photons are both
produced in the QGP phase and in the hadron-gas phase.  The production
of direct photons per unit time and volume for a given temperature has
been calculated for a QGP and for a hadron gas
\cite{Arnold:2001ms,Turbide:2003si}.  These state-of-the-art results
indicate that at the same temperature the photon production rates in a
QGP and a hadron gas are very similar. The photon rates must be
convoluted with the space-time evolution of the fireball in order to
obtain prediction for thermal photon production in nucleus-nucleus
collisions. Pure hadron gas scenarios and scenarios with phase
transition can be compared to data in order to make statements about
the formation of a QGP in these collisions.

Besides thermal and non-thermal direct photons a third category of
hard+thermal photons is conceivable in nucleus-nucleus collisions.  It
was suggested that a significant number of direct photons results from
processes in which a fast parton from initial hard scattering
interacts with a thermalized parton of the QGP \cite{Fries:2002kt}.
These processes include Compton scattering
($\mathrm{q_\mathrm{hard}}+\mathrm{g}_\mathrm{QGP} \rightarrow \gamma
+ \mathrm{q}$) and annihilation
($\mathrm{q_\mathrm{hard}}+\bar{\mathrm{q}}_\mathrm{QGP} \rightarrow
\gamma + \mathrm{g}$).  Recently, it was pointed out that also induced
photon bremsstrahlung from multiple scattering of a fast quark in a
QGP might be a significant source of direct photons
\cite{Zakharov:2004bi}.  The process is similar to medium induced
gluon emission which is the dominant energy loss mechanism for fast
quarks in a QGP.

\section{Measurement of Direct Photons}
\label{sec:meas}
The direct-photon measurements presented here were made with highly
segmented electro-magnetic calorimeters. In low-multiplicity
environments, as {\it e.g.} p+p or $\mathrm{p}+\bar{\mathrm{p}}$
collisions, one can considerably reduce the decay photon background by
accepting only those hits as direct-photon candidates which don't form
an invariant mass in the mass range of the $\pi^0$ or $\eta$ meson
with other photon hits in the same event. Moreover, direct photons
produced in these collisions via a direct production mechanism, whose
momentum is therefore balanced by a particle jet on the away side, can
be identified by so-called isolation cuts which set a limit on the sum
of the energies of particles in a cone around the photon.  Due to the
high multiplicity of produced particles it is difficult to apply such
cuts in central collisions of heavy nuclei.

The first step in the measurement direct photons in nucleus-nucleus
collisions is the measurement of the inclusive photon $p_\mathrm{T}$
spectrum, {\it e.g.}  of the spectrum that includes direct photons as
well as decay photons. The fraction of charged particle background
hits is usually measured and subtracted with the aid of position
sensitive detectors located directly in front of the calorimeter. A
correction for the contribution of fake photon hits produced by
neutrons and antineutrons, which cannot be determined experimentally,
is obtained on the basis of detailed Monte-Carlo simulations. The
second step is an accurate measurement of the $p_\mathrm{T}$ spectrum
of neutral pions and $\eta$ mesons with the same detector. Based on
these spectra the expected decay photons are calculated. This is
usually done with Monte-Carlo calculations which also take into
account the small contribution from other hadrons ({\it e.g.} $\eta'$,
$\omega$) with a decay branch into photons. Since the
high-$p_\mathrm{T}$ part of the $\pi^0$ spectrum in a nucleus-nucleus
collision can be described by a power-law and the dominant
contribution to the decay-photon background comes from $\pi^0$ decays,
the following formula is a useful estimate of the background photons
per $\pi^0$:
\begin{equation}
  \frac{1}{p_\mathrm{T}} \frac{\mathrm{d}N_{\pi^0}}{\mathrm{d}p_\mathrm{T}} 
  \propto p_\mathrm{T}^{-n}
  \; \Rightarrow \; 
  \frac{\gamma_{\pi^0}^\mathrm{decay}}{\pi^0} \approx \frac{2}{n-1}.
\end{equation}
At RHIC energies the high-$p_\mathrm{T}$ $\pi^0$ spectrum can be
described with $n \approx 8$, so that at a given $p_\mathrm{T}$ one
roughly expects $\gamma_{\pi^0}^\mathrm{decay}/\pi^0 \approx 0.28$
decay photons per neutral pion.

The final step in the determination of the direct-photon spectrum is
the subtraction of the calculated decay photon spectrum from the
inclusive photon spectrum. The invariant direct-photon yield
$\gamma_\mathrm{direct} \equiv E\,\mathrm{d}^3N/\mathrm{d}^3p$ is
calculated as a fraction of the inclusive photon spectrum:
\begin{eqnarray}
\gamma_\mathrm{direct} & = & \gamma_\mathrm{incl} - \gamma_\mathrm{decay}
  = (1 - \frac{\gamma_\mathrm{decay}}{\gamma_\mathrm{incl}}) 
    \cdot \gamma_\mathrm{incl} \\
 & \equiv & (1 - R_\gamma^{-1}) \cdot \gamma_\mathrm{incl}, 
\end{eqnarray}
where  
\begin{equation}
\label{eq:rgamma}
R_\gamma = \frac{\gamma_\mathrm{incl}}{\gamma_\mathrm{decay}}
         \approx \frac{(\gamma_\mathrm{incl}/\pi^0)_\mathrm{meas}}
                {(\gamma_\mathrm{decay}/\pi^0)_\mathrm{calc}}.
\end{equation}
The double ratio as given by the rightmost expression in
Eq.~\ref{eq:rgamma} is used to calculate $R_\gamma$ because in the
ratio $(\gamma_\mathrm{incl}/\pi^0)_\mathrm{meas}$ correlated
systematic uncertainties of the photon and $\pi^0$ measurement
partially cancel. The systematic uncertainties related to the energy
scale of the detector and to corrections which take detector effects
like energy smearing and shower overlap into account are examples for
such correlated systematic uncertainties. The ratio
$(\gamma_\mathrm{decay}/\pi^0)_\mathrm{calc}$ is the result of the
(Monte-Carlo) calculation of the expected background decay photons
from $\pi^0$, $\eta$ and other hadron decays per neutral pion.
$R_\gamma$ contains the entire statistical and systematic significance
of the direct-photon signal. In the calculation of the respective
direct-photon $p_\mathrm{T}$ spectrum only those systematic
uncertainties that canceled in the double ratio must be added.

\section{CERN SPS Results}
\label{sec:sps}
Early attempts to measure a direct-photon signal in fixed-target
experiments with proton, oxygen, and sulfur beams at a beam energy of
$200\,$GeV per nucleon at the CERN SPS resulted in upper limits on the
direct-photon signal.  At moderate transverse momenta ($p_\mathrm{T}
\gtrsim 0.5\,$GeV/$c$) upper limits on
($\gamma_\mathrm{direct}/\gamma_\mathrm{decay}$) on the order of 15\%
were determined
\cite{Akesson:1989tk,Albrecht:1990jq,Baur:1995gt,Albrecht:1995fs}.
The WA80 experiment found an (non-significant) average direct-photon
excess in central $^{32}$S+Au collisions in the range $0.5\,$GeV/$c
\le p_\mathrm{T} \le 2.5\,$GeV/$c$ of $5.0\% \pm 0.8\%(\mathrm{stat})
\pm 5.8\%(\mathrm{syst})$.  Comparisons with theoretical calculations
showed that the respective $p_\mathrm{T}$ dependent upper limits were
consistent with scenarios with a phase transition as well as with
scenarios without a phase transition \cite{Gale:2003iz}.

\begin{figure}
  \centerline{
    \includegraphics[width=0.45\textwidth]{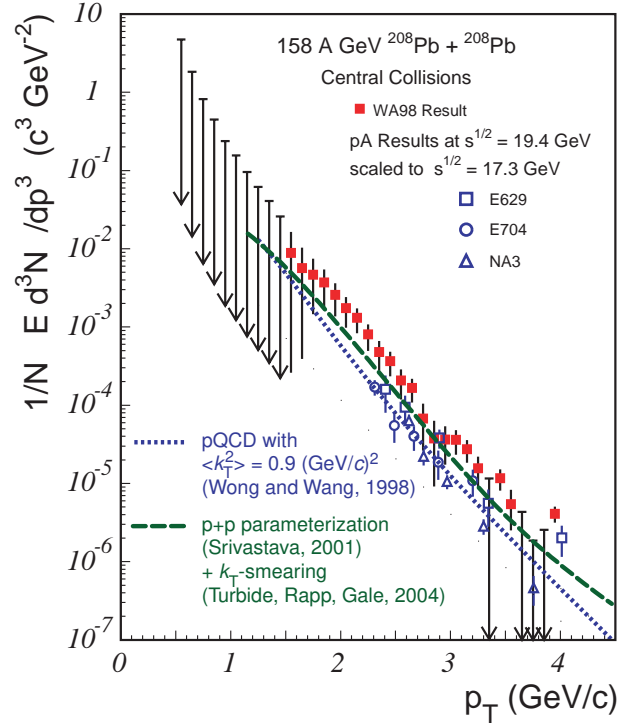}}
  \caption{The invariant direct-photon multiplicity as a function of
    the transverse momentum $p_\mathrm{T}$ in central Pb+Pb collisions
    at $\sqrt{s_\mathrm{NN}} = 17.2\,$GeV.  The central event class
    corresponds to $0-12.7\%$ of the WA98 minimum bias trigger which
    in turn corresponds roughly to $0-11\%$ of the inelastic Pb+Pb
    cross section. The error bars indicate combined statistical and
    systematic uncertainties. The upper edges of the arrows represent
    $90\,\%$ C.L. upper limits on the direct-photon yield. The WA98
    data points are compared with scaled p+p and p+C results, pQCD
    calculation which include $k_\mathrm{T}$ broadening, and scaled
    parameterizations of direct-photon yields in p+p collisions.}
  \label{fig:wa98_spec} 
\end{figure}

The first significant direct-photon signal in ultra-relativistic
nucleus-nucleus collisions was found at the CERN SPS by the
fixed-target experiment WA98 in Pb+Pb collisions at a beam energy of
$158\,$GeV per nucleon corresponding to a center-of-mass energy of
$\sqrt{s_\mathrm{NN}} = 17.2\,$GeV per nucleon-nucleon pair
\cite{Aggarwal:2000th}. A direct-photon signal of about
$\gamma_\mathrm{direct}/\gamma_\mathrm{decay} \approx 20\% \pm
9\%(\mathrm{syst})$ in the range $p_\mathrm{T} \gtrsim 2\,$GeV/$c$ was
found in central Pb+Pb collisions.  No significant signal was observed
in peripheral collisions. The extracted invariant direct-photon yield
for central Pb+Pb collisions is shown in Figure~\ref{fig:wa98_spec}.

The obvious question is whether the direct-photon spectrum in central
Pb+Pb collisions can be explained simply by hard scattering processes
or whether an additional thermal component is necessary. A first step
towards answering this question is the comparison of the WA98 results
with p+p and p+C results scaled by the average number of inelastic
nucleon-nucleon collision ($\langle N_\mathrm{coll}\rangle \approx
660$) for the central Pb+Pb class. A systematic uncertainty results
from the fact that the p+p and p+C data were measured at
$\sqrt{s_\mathrm{NN}}=19.4\,$GeV and therefore need to be scaled down
to $\sqrt{s_\mathrm{NN}}=17.2\,$GeV assuming scaling in $x_\mathrm{T}
= 2 p_\mathrm{T}/\sqrt{s_\mathrm{NN}}$. Figure~\ref{fig:wa98_spec}
shows that the scaled p+p and p+C results lie systematically below the
Pb+Pb data which suggests nuclear effects beyond the simple
$N_\mathrm{coll}$ scaling of the direct-photon yields. Since no p+A
data are available for $p_\mathrm{T} < 2.5\,$GeV/$c$ the Pb+Pb results
are compared to a p+p pQCD calculation \cite{Wong:1998pq} and to a
parameterization of direct-photon yields in p+p collisions
\cite{Turbide:2003si}, both scaled by $N_\mathrm{coll}$.  With a
transverse momentum broadening of $\langle k_\mathrm{T}^2 \rangle =
0.9\,$GeV$^2$ the pQCD calculation offers a good description of the
p+A data. By contrast, the Pb+Pb data for $p_\mathrm{T} \lesssim
2.5\,$GeV/$c$ are noticeably above the pQCD expectation.  The
parameterization of the direct-photon yields shown in
Figure~\ref{fig:wa98_spec} is the parameterization from
\cite{Srivastava:2001bw}, modified according to the additional
$k_\mathrm{T}$ broadening expected in nuclear targets (Cronin effect)
\cite{Turbide:2003si}. This parameterization hardly gives a
satisfactory descriptions of the Pb+Pb data which hints at the
presence of thermal photon sources in Pb+Pb collisions.  The effect of
nuclear $k_\mathrm{T}$ broadening on the direct-photon yield was
systematically studied in \cite{Dumitru:2001jx}. The authors conclude
that regardless of the amount of $k_\mathrm{T}$ broadening that is
assumed in their calculations the Pb+Pb data in the range
$p_\mathrm{T} \lesssim 2.5\,$GeV/$c$ cannot be described by
leading-order pQCD. Thus, it appears unlikely that the WA98 data can
be described by hard scattering processes alone. Firmer conclusions,
however, can only be drawn with better p+p and p+A reference data.

\begin{figure}
  \centerline{
    \includegraphics[width=0.45\textwidth]{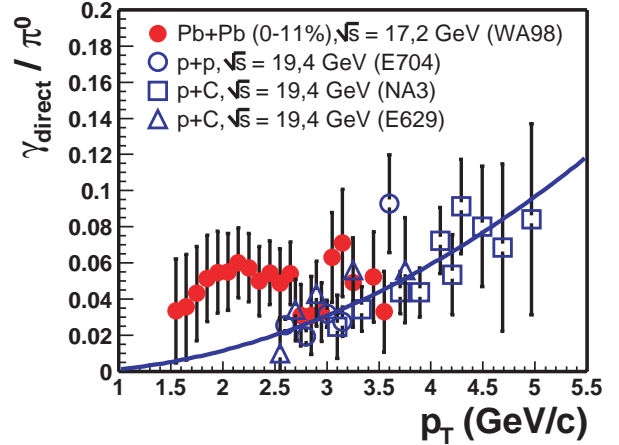}}
  \caption{Invariant direct-photon multiplicity divided by the
    invariant neutral pion multiplicity for p+A and central Pb+Pb
    collisions as a function of transverse momentum. No $x_\mathrm{T}$
    scaling of the p+A yields was applied here. The solid line 
    represents a fit to the p+A data with a functional form which can
    well describe the $\gamma_\mathrm{direct}/\pi^0$ obtained in pQCD
    calculations.}
  \label{fig:wa98_gam_pi0_ratio} 
\end{figure}

A striking feature of the WA98 data becomes visible in
Figure~\ref{fig:wa98_gam_pi0_ratio} which shows the
$\gamma_\mathrm{direct}/\pi^0$ ratio as a function of $p_\mathrm{T}$
for central Pb+Pb collisions along with the p+p and p+C reference
data. For $p_\mathrm{T} \gtrsim 2.5\,$GeV/$c$ the Pb+Pb and p+A
results are in agreement. The solid line represents a fit to the p+A
data with a functional form which can well describe the
$\gamma_\mathrm{direct}/\pi^0$ obtained in pQCD calculations.  In the
range $1.5\,$GeV/$c < p_\mathrm{T} < 2.5\,$GeV/$c$ the Pb+Pb data
exhibit a characteristic deviation from the solid line which reflects
the expected behavior for direct photon and neutral pion production
in hard scattering processes. A possible explanation for this
observation is the production of thermal direct photons in central
Pb+Pb collisions at the CERN SPS.

\begin{figure}
  \centerline{
    \includegraphics[width=0.45\textwidth]{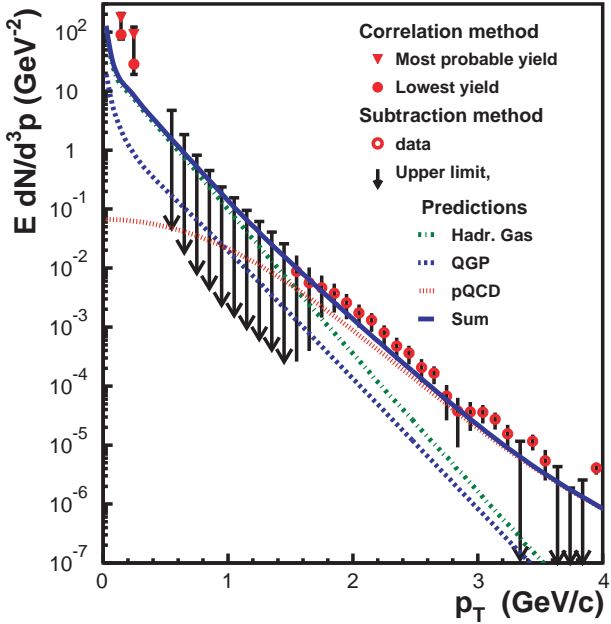}}
  \caption{Comparison of the WA98 data with a model calculation
    \cite{Turbide:2003si} which includes a phase transition from a QGP
    to a hadron gas. The data points at low $p_\mathrm{T}$ obtained
    from Hanbury Brown-Twiss interferometry cannot be explained by
    this calculation.}
  \label{fig:wa98_new} 
\end{figure}

Several authors have calculated the expected direct-photon spectrum in
central Pb+Pb collisions at the CERN SPS in scenarios in which thermal
photons are produced in a QGP and in a hadron gas in addition to
prompt direct photons. In some of these models the evolution of the
reaction zone is determined with hydrodynamic calculations, other
models employ simple parameterizations of the fireball evolution. One
objective of these model comparisons is the determination of the
initial temperature of the fireball directly after thermalization. In
\cite{Turbide:2003si} it was shown that in the absence of additional
nuclear $k_\mathrm{T}$ broadening ({\it i.e.} in the absence of the
Cronin effect) the WA98 data can be described with an initial
temperature of $T_\mathrm{i} = 270\,$MeV in a scenario with phase
transition from a QGP to a hadron gas. However, if nuclear
$k_\mathrm{T}$ broadening is assumed ($\langle \Delta k_\mathrm{T}^2
\rangle = 0.2\,$GeV$^2$) the data can be well described with a much
lower initial temperature of $T_\mathrm{i} = 205\,$MeV.  For the
latter case the different contributions to the calculated spectrum and
the comparison of the sum of all contributions with the WA98 data are
shown in Figure~\ref{fig:wa98_new}.

According to \cite{Srivastava:2004xp} the WA98 data suggest a very
small formation time $\tau_0 \approx 0.2\,$fm/$c$ and a high initial
temperature on the order of $T_\mathrm{i} \approx 335\,$MeV. In
\cite{Renk:2003fn} the WA98 data are found to be consistent with
initial temperatures in the range $250\,$MeV $\lesssim T_\mathrm{i}
\lesssim 370\,$MeV. In \cite{Huovinen:2001wx} the data are compared
with hydrodynamic calculations with and without a QGP phase.  In both
scenarios the authors can describe the data with initial temperatures
between $210\,$MeV $\lesssim T_\mathrm{i} \lesssim 260\,$MeV.  Thus,
one can conclude that the WA98 data can be naturally explained in a
QGP scenario but don't offer a direct proof for the existence of this
state of matter. Unfortunately there are large variations in the
extracted initial temperatures but one should also note that most
models suggest initial temperatures $T_\mathrm{i}$ above the critical
temperature $T_\mathrm{c}$ for the QGP phase transition.

The method of measuring direct photons by subtracting the expected
decay-photon yield from the measured inclusive yield depends on the
accurate measurement of the $\pi^0$ $p_\mathrm{T}$ spectrum. Largely
due to systematic uncertainties of the $\pi^0$ measurement at low
$p_\mathrm{T}$ no significant direct-photon signal could be observed
with the subtraction method in central Pb+Pb collisions for
$p_\mathrm{T} < 1.5\,$GeV/$c$. An alternative method of extracting the
direct-photon yield is based on Bose-Einstein two-particle
correlations of direct photons at small relative momenta which is
absent for photons from hadron decays. This method requires a very
good understanding of background effects which could mimic two-photon
correlations. These effects include splitting of single particle
showers into two nearby clusters, misidentification of
$e^{+}e^{-}$-pairs from photon conversion as photon hits, and
Bose-Einstein correlations of charged pions misidentified as photons.
In addition to the expected background effects the WA98 experiment
observed genuine two-photon correlations which were attributed to
Bose-Einstein correlations \cite{Aggarwal:2003zy}. From the observed
correlation strength it was possible to determine the direct-photon
yield in the range $0.1\,$GeV/$c < p_\mathrm{T} < 0.3\,$GeV/$c$. The
new data points are compared to a theoretical calculation in
Figure~\ref{fig:wa98_new}. The measured direct-photon yields exceed
this model calculation which attributes the dominant contribution in
this $p_\mathrm{T}$ range to direct photons from the hadron gas phase.

\section{RHIC Results}
\label{sec:rhic}
\begin{figure}
  \centerline{
    \includegraphics[width=0.45\textwidth]{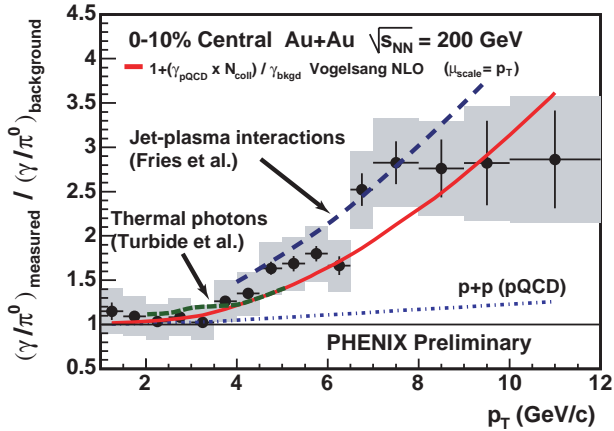}}
  \caption{Ratio of the measured inclusive photon spectrum and the
    expected photon spectrum from the decay of hadrons ($\pi^0$,
    $\eta$, etc.) for central Au+Au collisions at
    $\sqrt{s_\mathrm{NN}}= 200\,$GeV. A clear direct-photon signal is
    visible for $p_\mathrm{T} \gtrsim 4\,$GeV/$c$. The solid line
    indicates the expected direct-photon excess for the case that the
    direct photon yield in the Au+Au collisions is given by the
    direct-photon yield in p+p collisions (at the same energy
    $\sqrt{s_\mathrm{NN}}$) scaled by the average number $\langle
    N_\mathrm{coll}\rangle$ of binary nucleon-nucleon collisions for
    the given Au+Au centrality class. The long dashed line represents
    the predicted direct-photon signal for a calculation that includes
    photon production due to the interaction of partons from hard
    scattering with thermalized partons of the QGP (jet-plasma
    interactions) \cite{Fries:2002kt}.  For $2\,\mathrm{GeV}/c <
    p_\mathrm{T} < 5\,\mathrm{GeV}/c$ an estimate of the expected
    thermal photon signal is compared to the data. }
  \label{fig:phenix_photons} 
\end{figure}

By comparing the measured inclusive photon $p_\mathrm{T}$ spectrum
with the expected photons from hadron decays the PHENIX experiment was
able to observe a direct-photon signal in Au+Au collisions at
$\sqrt{s_\mathrm{NN}} = 200\,$GeV up to transverse momenta of
$p_\mathrm{T} \approx 12\,$GeV/$c$ \cite{Frantz:2004gg}. PHENIX
employed two different types of highly segmented electro-magnetic
calorimeters: a lead-scintillator sandwich calorimeter (PbSc) and a
leadglass Cherenkov (PbGl) calorimeter.  The independent direct-photon
measurements made with these two detectors, which have a different
response to hadrons, were found to be in agreement.  The PbSc and PbGl
results were averaged in order to reduce the statistical uncertainty
of the direct-photon signal. Figure~\ref{fig:phenix_photons} shows the
ratio $R_\gamma$ of the measured inclusive photons to the expected
decay photons in central Au+Au collisions. A significant direct-photon
signal is visible for $p_\mathrm{T} \gtrsim 4\,$GeV/$c$. Above
$p_\mathrm{T} \approx 7\,$GeV/$c$ the number of direct photons exceeds
the number of photons from hadron decays.

In the $p_\mathrm{T}$ range where the photon excess is observed direct
photons are expected to result from initial hard parton-parton
scatterings which can be described by pQCD. Neutral pions and charged
hadrons at high $p_\mathrm{T}$ in central Au+Au collisions at RHIC
were found to be suppressed relative to the cross section in p+p
collisions scaled by a geometrical factor $\langle T_\mathrm{AB}
\rangle_\mathrm{f} \equiv \langle N_\mathrm{coll} \rangle /
\sigma_\mathrm{inel}^\mathrm{p+p}$ \cite{Adcox:2004mh}. This
$N_\mathrm{coll}$ scaling is only expected for hard-scattering
processes. The high-$p_\mathrm{T}$ hadron suppression can be explained
by energy loss of fast partons from initial hard scatterings in the
medium of high color charge density produced in Au+Au collisions.
Thus, in this picture the hadron suppression is attributed to a final
state effect. Alternative models suggested that the hadron suppression
is related to an initial state effect, namely to special properties of
the gluon densities in the ground state Au nucleus
\cite{Kharzeev:2002pc}.  High-$p_\mathrm{T}$ direct photon and {\it
e.g.} neutral pion production in Au+Au collisions are sensitive to the
same initial flux of incoming partons. High-$p_\mathrm{T}$ direct
photons, however, are largely unaffected by the created hot
and dense medium.

In Figure~\ref{fig:phenix_photons} the measured direct-photon signal
is compared to the expected signal in case $N_\mathrm{coll}$ scaling
of the direct-photon multiplicities holds for Au+Au collisions.  The
solid line represents the ratio
\begin{equation}
  \frac{\gamma_\mathrm{incl}^\mathrm{Au+Au}}
  {\gamma_\mathrm{decay}^\mathrm{Au+Au}}
  = 
  \frac{\gamma_\mathrm{direct}^\mathrm{Au+Au} 
    + \gamma_\mathrm{decay}^\mathrm{Au+Au}}
  {\gamma_\mathrm{decay}^\mathrm{Au+Au}}
  =
  1 + \frac{\langle N_\mathrm{coll} \rangle 
    \cdot \gamma_\mathrm{direct}^\mathrm{p+p}}
  {\gamma_\mathrm{decay}^\mathrm{Au+Au}}.
\end{equation}
The result of a NLO pQCD calculation
\cite{Gordon:1993qc,Gordon:1994ut} was used as p+p direct-photon
reference ($\gamma_\mathrm{direct}^\mathrm{p+p}$).  Comparisons with
final direct-photon data for p+p collisions at $\sqrt{s} = 200\,$GeV
from RHIC Run-2 \cite{Adler:2005qk} and preliminary results from Run-3
\cite{Okada:2005in} justify the use of the pQCD calculation as a
reference. The intrinsic uncertainty of the pQCD photon yield
($\approx 20\%$) due to the choice of the QCD scales is not shown. The
agreement of the solid line in Figure~\ref{fig:phenix_photons} with
the data point shows that in central Au+Au collisions at RHIC
high-$p_\mathrm{T}$ direct photons are not suppressed and follow
$N_\mathrm{coll}$ scaling. The same behavior was also found for less
central Au+Au collisions \cite{Frantz:2004gg}. This shows that the
observed suppression of hadrons at high $p_\mathrm{T}$ is due to final
state effects and therefore supports the interpretation of the hadron
suppression as the result of jet-quenching.

According to the calculation in \cite{Turbide:2003si} thermal direct
photons from a QGP produced in central Au+Au collisions at RHIC might
be the dominant direct-photon source in the range $1\,$GeV/$c \lesssim
p_\mathrm{T} \lesssim 3\,$GeV/$c$.  In Figure~\ref{fig:phenix_photons}
the expected thermal-photon signal is indicated. This expectation was
obtained by adding the ratio of thermal photons (from the QGP and the
hadron gas phase) and pQCD photons from \cite{Turbide:2003si} to the
pQCD curve (solid line) shown in Figure~\ref{fig:phenix_photons}.
This estimate illustrates that the expected signal from thermal direct
photons is rather small (on the order of 10\,\%). With the current
systematic uncertainties no statement can be made about the presence
of thermal direct photons.

Figure~\ref{fig:phenix_photons} furthermore shows an estimate of the
direct-photon signal in central Au+Au collisions at RHIC in case
jet-plasma interactions as calculated in \cite{Fries:2002kt}
contribute to the direct-photon production (long dashed line). The
ratio of photons from the interaction of fast quarks from hard
scattering with the QGP and photons from (leading-order) pQCD
determined in \cite{Fries:2002kt} was added to the (next-to-leading
order) pQCD curve (solid line) in Figure~\ref{fig:phenix_photons}. The
leading order photon calculation in \cite{Fries:2002kt} doesn't
include a $K$ factor which can be used to account for higher order
contributions. In \cite{Dumitru:2001jx} the leading order direct
photon spectra were {\it e.g.} multiplied by a factor $K = 2$.
Therefore, the result in the presence of jet-plasma interactions is
only a rough estimate. It shows, however, that with the uncertainties
of the preliminary PHENIX data it is not possible to decide whether
there's a strong contribution of photons from jet-plasma interactions.
In non-central nucleus-nucleus collisions the path length of a parton
from hard scattering depends on the angle with respect to the reaction
plane and is longest for the direction perpendicular to the reaction
plane.  Hence, direct photons from jet-plasma interactions should be
predominantly produced perpendicular to the reaction plane. This
characteristic property offers an experimental handle on the detection
of this photon source.

\section{Summary}
\label{sec:summary}
Direct photons in ultra-relativistic collisions of heavy nuclei were
observed by the WA98 experiment at the CERN SPS and by the PHENIX
experiment at RHIC. The WA98 data can naturally be explained in a QGP
scenario. However, also models which don't assume a phase transition
to a QGP can describe the data and cannot be ruled out.  PHENIX
measured direct photons in the high-$p_\mathrm{T}$ regime where hard
scattering is expected to be the dominant production mechanism. Unlike
high-$p_\mathrm{T}$ hadrons direct photons are not suppressed in
central Au+Au collisions at RHIC.  This shows that the suppression of
high-$p_\mathrm{T}$ hadrons is a final state effect related to
properties of the hot and dense medium created in central Au+Au
collisions at RHIC.

%
\bibliographystyle{epjckr}
\bibliography{hp04_photons}

\end{document}